\documentclass[apj]{emulateapj}
\usepackage[colorlinks,linkcolor={blue},citecolor={blue},urlcolor={red}]{hyperref}
\usepackage{epsfig}
\usepackage{graphicx}
\usepackage{natbib}
\usepackage{amsmath}
\usepackage{amsfonts}
\usepackage{amssymb}
\usepackage{xcolor}
\def\mpchi{\,h^{-1}{\rm {Mpc}}}

\def\kms{\,{\rm {km\, s^{-1}}}}
\def\msun{{\rm M_\odot}}

\begin{document}

\title{The distribution of faint satellites around central galaxies in 
the CFHT Legacy Survey}
\author{C. Y. JIANG, Y. P. JING, CHENG LI}
\affil{Key Laboratory for Research in Galaxies and Cosmology of Chinese 
Academy of Sciences, Shanghai Astronomical Observatory, Nandan Road 80, 
Shanghai 200030, China}

\begin{abstract}

We investigate the radial number density profile and the abundance
distribution of faint satellites around central galaxies in the low 
redshift universe using the CFHT Legacy Survey. We consider three 
samples of central galaxies with magnitudes of $M_{\rm r}=-21$, $-22$, 
and $-23$ selected from the Sloan Digital Sky Survey (SDSS) group 
catalog of Yang et al.. The satellite distribution around these central 
galaxies is obtained by cross-correlating these galaxies with the 
photometric catalogue of the CFHT Legacy Survey. The projected radial 
number density of the satellites obeys a power law form with the 
best-fit logarithmic slope of $-1.05$, independent of both the central 
galaxy luminosity and the satellite luminosity. The projected cross 
correlation function between central and satellite galaxies exhibits
a non-monotonic trend with satellite luminosity. It is most 
pronounced for central galaxies with $M_{\rm r}=-21$, where the 
decreasing trend of clustering amplitude with satellite 
luminosity is reversed when satellites are fainter than central 
galaxies by more than 2 magnitudes. A comparison with the satellite 
luminosity functions in the Milky Way and M31 shows that the Milky 
Way/M31 system has about twice as many satellites as around a typical central 
galaxy of similar luminosity. The implications for theoretical 
models are briefly discussed.

\end{abstract}


\section{Introduction}

The distribution of satellite galaxies around the central galaxy in a
dark matter halo carries important information for the underlying 
cosmology, dark matter properties and galaxy formation processes. 
It is specified by the abundance distribution of satellites with 
varying properties (luminosity, color, star formation rate and so on), 
and the spatial distribution of the satellites. A clear picture of how 
satellite galaxies are distributed in the real universe provides a critical 
test for the theoretical models of cosmology and galaxy formation.

Current N-body studies have shown that the abundance of subhalos that
satellites inhabit is nearly scale invariant, if the subhalo mass
scaled by the virial mass of the host halo is considered
\citep{madau08,angulo09,han11}. However, due to the complex baryonic
physics involved in galaxy formation, the relation between the
dark matter mass of a halo and its central galaxy mass is not linear 
\citep[e.g.][]{yang03,tinker05,mandelbaum06,wanglan06,yang07,zheng07,
guoqi10,moster10,wanglan10,more11,li12}.  Furthermore, when a galaxy
together with its surrounding halo infalls into a bigger halo and
becomes a satellite, physical processes such as tidal stripping, ram
pressure and gas starvation begin to take effects, which further
complicates the galaxy-subhalo relation. Thus, the satellite
distribution can not be inferred from the subhalo distribution
nontrivially, and observations of the satellite distribution are expected to
provide important clues to these physical processes.

From rich clusters to Milky Way (MW)-sized groups, the luminosity of
central galaxies changes by $\sim 1$ order of magnitude, while the
mass of their host halos spans $\sim 3$ orders of magnitude. The
luminosity-halo mass relation flattens at the massive halo end, as a
result of lower efficiency of converting baryonic matter to light in
more massive halos.  We then expect that the number of satellites with
fixed magnitude difference between the satellite and the central
galaxy ($\Delta m=M^{s}-M^{c}$) is larger for more luminous central
galaxies, unlike the nearly scale-independent behavior in the subhalo
population.

The Local Group satellites, due to their proximity, have been observed
in the most detail, and the observational data has been used to test
and calibrate various galaxy formation models.  The apparent tension
between the large number of subhalos in $\Lambda$CDM simulations and
only dozens of satellites observed in the Local Group has challenged
theorists, making them either scrutinize various baryonic processes that
might suppress star formation in small halos
\citep{kravtsov04,koposov09}, or investigate the possibility of
alternative dark matter models \citep{zavala09,lovell12}.

Despite possible underpopulation of the satellites as a whole in the
Milky Way compared with theoretical subhalos, some authors have found
that the existence of the two Magellanic Clouds (MCs) seems incompatible
with the $\Lambda$CDM universe.  \cite{koposov09} found that it is
very difficult to reproduce the Magellanic Clouds in models that can
solve the 'missing satellite' problem.  Under the assumption of
abundance matching, \cite{boylan-kolchin10} stated that there is only
a $10\%$ chance of finding two galaxies as bright as Magellanic Clouds
in a halo of virial mass $M_{\rm vir}=10^{12}\msun$.
\cite{busha11} reached a similar conclusion also with an abundance
matching method. They showed that, even before the abundance matching
is applied, the probability of hosting two satellites with maximum
circular velocity exceeding $v_{\max}=50\kms$ which is the lower bound
of Magellanic Clouds \citep{vanderMarel02,stanimirovic04} 
is only $\sim 8\%$ for halos of $M_{\rm vir}=1.2 \times 10^{12}\msun$ in
their simulation. This probability increases to $\sim 20\%$ for halos
of $M_{\rm vir}=2.6 \times 10^{12}\msun$.  However, it is
contradictory to the result of \cite{wang12}, in which the number of
subhalos having $v_{\rm max}>60\kms$ in halos with $M_{\rm vir}=2.6
\times 10^{12}\msun$ is already over $1$. They found that the number
of subhalos above the maximum circular velocity threshold can be
underestimated if the mass resolution of simulation is not high
enough.  The fact that the number of particles within the virial
radius in \cite{busha11} is smaller than the minimum number needed to
achieve convergence as listed in \cite{wang12} may account for part of
the difference in the probability of finding two subhalos like
Magellanic Clouds. 

On the other hand, it is doubted that the Milky Way, as a single case,
may be representative of galaxies with a comparable
luminosity. Some recent works have been dedicated to statistical
measurements of satellite abundance in large scale
surveys. \cite{guo11} studied the luminosity function of satellites
around isolated primaries in the Sloan Digital Sky Survey 
\citep[SDSS;][]{york00},
finding that the number of satellites brighter than $M_{\rm v}=-14$ in
MW/M31-like galaxies is lower than that in MW/M31 by a factor of
2. \cite{cosmos} used imaging and photometric redshift catalogs from 
the COSMOS survey to model the satellite distributions. Their accumulated
luminosity function of satellites in MW-luminosity hosts is consistent
with that in \cite{guo11}. \cite{robotham12} searched for the 
Milky Way-Magellanic Clouds analogues in the Galaxy and Mass Assembly (GAMA) survey,
finding that only $3.4\%$ of MW mass galaxies have two companions at 
least as massive as the small Magellanic Cloud, and only $0.4\%$ have
an analogous MW-MCs system if these galaxies are all restricted to be
late-type and star-forming.

The spatial distribution of satellite galaxies builds upon that of
subhalos. Many physical processes that affect the luminosity-halo mass
relation are dependent on the cluster/group-centric radius. Therefore,
the radial distribution of satellite galaxies provides indispensable
information of various processes that have shaped the luminosity-halo
mass relation.  $\Lambda$CDM simulations have confirmed that subhalos
are distributed less centrally concentrated in the host halo than the
dark matter, exhibiting a central core in the radial distribution of
their number density \citep{diemand04,gao04,ludlow09,gao12}. In addition, the
number density profile of subhalos has been shown to have a similar 
shape over a wide mass range.  The radial distribution of satellite 
galaxies, on the other hand, is usually described by a power law, 
$\Sigma(r_{\rm p}) \sim {r_{\rm p}}^{\rm \alpha}$, where $\Sigma(r_{\rm p})$ 
is the projected number density of satellites measured in observations.
However, no consensus has been reached regarding the power law index
$\alpha$ which spans a wide range from -0.5 to -1.7
\citep{madore04,smith04,sales05,chen06}. Recent work has taken
advantage of the large data sample from surveys like SDSS and COSMOS
to study the satellite radial distribution. \cite{chen06} used the
SDSS spectroscopic sample to study the projected radial distribution
of satellites around isolated $L^{\ast}$ galaxies, constraining the
power law slope to be $-1.7$. \cite{watson10} modeled the clustering
of luminous red galaxies in SDSS in the framework of halo occupation
distribution (HOD). They found that on projected scales $0.016\mpchi
\le r_{\rm p} \le 0.42\mpchi$, the radial density profile of luminous
red galaxies is close to an isothermal distribution. \cite{cosmos}
detected satellites up to eight magnitude fainter than their host
galaxies (with stellar mass above $10^{10.5}\msun$) by subtracting the
host light profile in the COSMOS survey. They constructed parametrized
models for the satellite spatial distribution, number of satellites
per host, and background galaxies, and then inferred the parameters
using Markov Chain Monte Carlo (MCMC) method. The power law slope was
found to be $-1.1$, independent of host stellar mass and satellite
luminosities.

Observational studies mentioned above have either used a spectroscopic sample,
or combined a spectroscopic sample for central galaxies with a
deeper photometric sample to probe fainter satellites.  In the latter
case, the background must be subtracted properly due to the
lack of redshift information. So far, several background subtraction
methods have been used. For example, \cite{lares11} applied a color
cut, $g-r<1$, to identify satellite galaxies, taking galaxies with colors
beyond this range as background galaxies for the redshift range they
considered. However, this method failed at $r_{\rm p}<100$ ${\rm
  {kpc}}$. \cite{guo11} counted the number of galaxies in an annuli
300 ${\rm {kpc}}<r_{\rm p}<600$ ${\rm {kpc}}$ to estimate the
background. \cite{cosmos} used a similar method to build a prior in
modeling the background. \cite{liu11} randomized the host positions on
the sky and searched for projected companions from the
background. They excluded galaxies with photometric redshift $z_{\rm
  p}>0.23$ in the whole calculation to lower the noise, and then
corrected for this photo-z loss and for the under-deduction of
background.

In this work, we use the halo-based group catalog of the SDSS
constructed by Yang et al. (2007) to optimize the selection of central
galaxies. We use a deep photometric catalog of the CFHT Legacy
Survey, which is $\sim$ 3 magnitude deeper than the SDSS, to provide
reliable information about the faint satellite galaxies around the
central ones. Furthermore, we adopt a novel method of background
subtraction that does not rely on models of the local environment or the
photometric redshift information. Thus we are able to produce a
reliable measurement of the satellite abundance and the radial density
distribution for central galaxies as bright as or brighter than the
Milky Way.

The remaining sections are arranged as follows. Our methodology
is described in section 2, along with the dataset and
sample selection.  Results of the radial distributions and the
abundance of satellites are presented in sections 3 $\&$
4. Conclusions are given in section 5.  Throughout this paper, we
assume a $\Lambda$CDM cosmology with $\Omega_{\rm m}=0.3$,
$\Omega_{\rm \Lambda}=0.7$, and $H_{\rm 0}=100h \kms {\rm Mpc}^{-1}$
with $h=0.7$. All magnitudes are given in the AB system.

\section{Data and Methods}

\subsection{Data}

The data used in this work is based on the photometric catalogue of
the CFHT Legacy Survey \citep{gwyn11}.  The wide fields of the survey
consist of 171 square degree pointings, which is about 145 ${\rm
  deg}^2$ when the masked areas are excluded. The star-galaxy
separation is done based on the Spectral Energy Distribution (SED)
fitting and on the source size, closely following the method adopted
by \cite{coupon09}. The survey adopts a 5-band photometric system
($u^{\ast},g^{\prime},r^{\prime},i^{\prime},z^{\prime}$), with a
limiting magnitude of $r^{\prime}=25.9$. To ensure a high photometric
accuracy, we select galaxies with $r^{\prime}<23.0$ to be the
photometric sample from which satellites are searched for. In what follows, 
we omit the prime when referring to the absolute magnitude in 
$r^{\prime}$ band which will be written as $M_{\rm r}$.

We build our sample of central galaxies using the group catalogue 
constructed by \cite{yang07}. Their extracted galaxy catalog is based 
on the NYU-VAGC \citep{blanton05} which is selected from the SDSS 
\citep{york00} data release 7 (DR7; \citealt{abazajian09}), and 
supplemented with additional redshifts (e.g. from 2dF) to compensate 
for the untargeted galaxies due to fiber collisions. The resulting 
number of galaxy groups identified is about $4.7 \times 10^5$, with 
spectroscopic redshifts $0.01<z<0.2$.  After removing groups near the 
edges of the survey (see \citealt{yang07} for details), we cross match 
the central galaxies of these groups with those in the overlapped area 
of the CFHTLS. The final matched sample has $3068$ central galaxies, 
of which $82\%$ are more luminous than $M_{\rm r}=-20.5$.

\subsection{Methodology}
\label{sec:method}

We aim to statistically measure the number of satellite galaxies around 
central galaxies within a projected distance ${r_{\rm p}}^{m}$. Given the
fact that the satellites are faint and thus usually do not have
reliable redshift information\footnote{The photometric redshift of
  faint galaxies is usually very inaccurate for large photometry
  errors}, the number count within a projected radius around the central 
galaxies is contributed by both their satellite galaxies and
the background galaxies\footnote{The foreground galaxies have the
  same effect as the background ones, so we do not distinguish them in
  the paper}. Fortunately, the background galaxies are uncorrelated
with the central galaxies being considered, and the contribution of
background galaxies can be statistically subtracted as shown in 
\cite{wenting11}.

Let us consider a central galaxy at redshift $z$.  To subtract the
background contribution, we assume all galaxies in the photometric
survey lie at the same redshift $z$ as the central galaxy. Background
galaxies at redshift $z_{\rm b}$ with the same properties (luminosity
and spectral energy distribution), would have the same magnitude
$M_{\rm r}$, when these galaxies are assumed to lie at redshift $z$.
Likewise, background galaxies at other redshifts may contribute to
$M_{\rm r}$ galaxies at the assumed redshift $z$.  Since these
background galaxies with $M_{\rm r}$ are not correlated {\it
  statistically} with the central one, we can then subtract the
background contribution for an ensemble of central galaxies by using
the mean density of $M_{\rm r}$ galaxies. Specifically, we generate a
random sample that has the same masked areas and boundaries as the
observed sample, and has random points $13$ times as many as the
observed sample has. We first count the total number of $M_{\rm r}$
galaxies that are at a projected distance of $r_{\rm p}$ away from the
central galaxy.  Next, we measure how many background galaxies there
would be in the random sample.  Finally, we subtract the background
contribution to obtain the surface number density of satellites
$\Sigma (r_{\rm p})$, after accounting for the fraction of masked area.
The number of satellite galaxies within ${r_{\rm p}}^{m}$ is then 
\begin{equation}
N(<{r_{\rm p}}^{m})=\int_0^{{r_{\rm p}}^{m}}\Sigma(r_{\rm p})2\pi r_{\rm p} d r_{\rm p}. 
\end{equation}
The projected correlation function $w(r_{\rm p})$ is related to 
$\Sigma(r_{\rm p})$ by $\Sigma(r_{\rm p})=n_{\rm g} w_{\rm p}(r_{\rm p})$, 
where $n_{\rm g}$ is the 3-dimentional number density 
of galaxies in consideration.

We use Le Phare, a photometric redshift computing code
\citep{arnouts99,ilbert06}, to calculate the absolute magnitude
$M_{\rm r}$ for each galaxy assuming all satellite galaxies are at
three fixed redshifts: $z_{\rm ref}=0.06,0.12$ and $0.24$. Since our
central galaxy sample spans the redshift range $0.01<z<0.2$, we obtain
$M_{\rm r}$ at redshift $z$ by
\begin{equation}
 M_{\rm r}(z) = M_{\rm r}(z_{\rm ref}) + (DM(z_{\rm ref})-DM(z)) + (kcor(z_{\rm ref})-kcor(z)),
\end{equation}
where $DM$ is distance modulus. We compute the difference in the
$k$-correction, $kcor(z_{\rm ref})-kcor(z)$, by linearly extrapolating the
$k$-correction between the two adjacent reference redshifts.

Galaxies are distributed in the form of clustering, with more luminous
galaxies being more strongly clustered. Galaxies of characteristic luminosity
$L^{\ast}$ show a correlation length of $\sim 5\mpchi$ \citep[see,
  e.g.][]{zehavi05,li06}, which is much larger than the virial
radius of their host dark matter halos. Therefore, the number excess of satellites around central galaxies 
after deducting a uniform background comprises two components: one from 
satellites within the virial radius, and one from the correlated 
nearby structures outside of the virial radius along the line of sight. 
The projected correlation function thus consists of two parts,
\begin{equation}
w_{\rm p} (r_{\rm p}) = w_{\rm p1} (r_{\rm p}) + w_{\rm p2} (r_{\rm p}), 
\end{equation}
where
\begin{equation}
\label{eq:wp1}
w_{\rm p1} (r_{\rm p}) = 2\int_{r_{\rm p}}^{r_{\rm vir}} rdr\xi(r)(r^2-{r_{\rm p}}^2)^{-1/2},
\end{equation}
and
\begin{equation}
w_{\rm p2} (r_{\rm p}) = 2\int_{r_{\rm vir}}^{\infty} rdr\xi(r)(r^2-{r_{\rm p}}^2)^{-1/2}.
\end{equation}
Here $\xi(r)$ is the real space correlation function
\citep{davis83}. If $\xi(r)$ follows the power law form $\xi(r)=(r/r_{\rm
  0})^{-\gamma}$, then the overall projected correlation function
becomes
\begin{equation}
\label{eq:wp}
w_{\rm p} (r_{\rm p}) = r_{\rm p}(\frac{r_{\rm p}}{r_{\rm 0}})^{-\gamma}\Gamma\bigg
(\frac{1}{2}\bigg)\Gamma\bigg(\frac{\gamma-1}{2}\bigg)/\Gamma\bigg(\frac{\gamma}{2}\bigg).
\end{equation}

The contribution of real satellites (galaxies within $r_{\rm vir}$)  
to the surface density at the projected distance $r_{\rm p}$ can thus be obtained by  
\begin{equation}
\Sigma_{\rm i}(r_{\rm p}) = \Sigma (r_{\rm p})  \frac{w_{\rm p1}(r_{\rm p})}{w_{\rm p}(r_{\rm p})},
\end{equation}
where $\frac{w_{\rm p1}(r_{\rm p})}{w_{\rm p}(r_{\rm p})}$ is a decreasing
function of $r_{\rm p}$ because of the projection effect.  Therefore,
the total number of real satellites within the projected distance ${r_{\rm p}}^{m}$
would be
\begin{equation}
\label{eq:Ni}
N_{\rm i}(<{r_{\rm p}}^{m}) = \int_{0}^{{r_{\rm p}}^{m}} \Sigma (r_{\rm p}) 
\frac{w_{\rm p1}(r_{\rm p})}{w_{\rm p}(r_{\rm p})}2\pi r_{\rm p} dr_{\rm p}.
\end{equation}

According to equation (\ref{eq:Ni}), we need to know the ratio of 
$\frac{w_{\rm p1}(r_{\rm p})}{w_{\rm p}(r_{\rm p})}$ to obtain 
the number of real satellites within the projected distance ${r_{\rm p}}^{m}$.
The ratio of $\frac{w_{\rm p1}(r_{\rm p})}{w_{\rm p}(r_{\rm p})}$, 
which can be calculated with equations (\ref{eq:wp1}) and (\ref{eq:wp}), 
is independent of the correlation length $r_{\rm 0}$, and depends 
only on the power law index $\gamma$. The value of $\gamma$ can be inferred from
the projected radial distribution profile of satellites. The details
are described in section \ref{sec:radial_dis}. There we find a power
law index of $-2.05$ for central galaxies in the three magnitude bins
we consider (${M_{r}}^c=-21,-22$, and $-23$), irrespective of
satellite luminosities.  With $\gamma=-2.05$, for central galaxies
with $M_{\rm r}=-21,-22$, and $-23$, the percentage $\frac{N_{i}(<{r_{\rm p}}^{m})}
{N(<{r_{\rm p}}^{m})}$ of real satellites within the projected distance 
${r_{\rm p}}^{m}=0.3$ Mpc is $66\%, 79\%$ and $88\%$ respectively. 
The number of real satellites within the projected virial radius 
$N_{\rm i}(<r_{\rm vir})$ is equivalent to the number of satellites 
within the virial radius, which is $66\%$ of $N(<r_{\rm vir})$.

\section{Radial distributions of satellites}

In this section we study the projected radial distribution of
satellites.  We first consider the projected number
density profile, and we fit it with a power law to infer the
logarithmic slope $\gamma$ of $\xi(r)$ in order to feed the the method
elaborated in section \ref{sec:method}. We then calculate the
projected two-point correlation function $w_{\rm p}(r_{\rm p})$ to
investigate how the correlation function changes with satellite
luminosity.

\subsection{Projected surface density profile}
\label{sec:radial_dis}

We consider three central samples with magnitudes in bins centered at
${M_{\rm r}}^{\rm c}=-21,-22$, and $-23$ with an interval of 1 magnitude.
In the SDSS DR7 group catalogue, host halo masses are estimated using
the ranking of the groups according to their characteristic group
luminosity or stellar mass. \cite{yang07} showed that, the
characteristic stellar mass is more tightly correlated with the halo
mass than the characteristic group luminosity. Therefore, we choose
the halo mass estimated from the 
stellar mass to calculate the virial radius. In the faintest
luminosity bin with ${M_{\rm r}}^{\rm c}=-21$, the halo masses are about
$90\%$ complete.  Since this luminosity is close to that of the Milky
Way, we consider Milky Way-sized host halos: 1.0
$\times10^{12}\msun<M_{\rm halo}<$3.0 $\times10^{12}\msun$
\citep[e.g.][]{wilkinson99,sakamoto03,dehnen06,xue08}, and draw the
virial radius from a cosmological N-body simulation. The simulation is
the same as that used in \cite{jiang08}, and the details can be found
there. The resulting virial radius is $0.3$ Mpc. Those of the two more
luminous bins are $0.43$ Mpc and $0.73$ Mpc respectively.

\begin{figure}
\begin{center}
\plotone{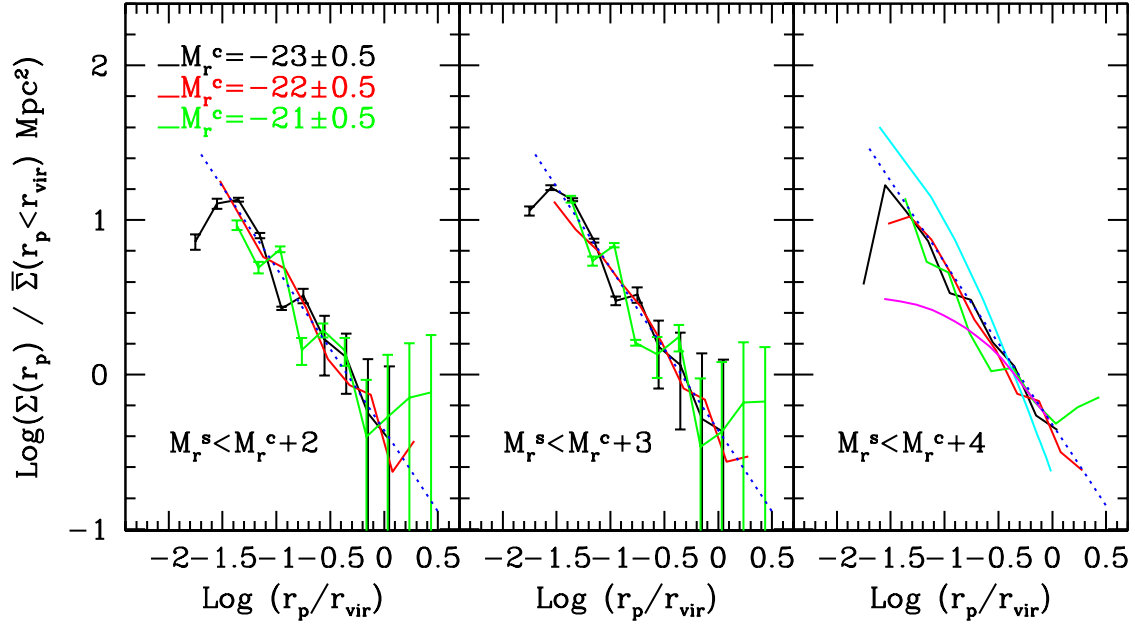}
\caption{\label{fig:radial_dis_rescl}The surface density of satellites 
as a function of the distance to the central galaxies. The surface 
density is scaled by the mean surface density within the virial radius, 
and the distance is scaled by the virial radius. From left to right: 
${M_{r}}^{s}<{M_{r}}^{c}+2, {M_{r}}^{s}<{M_{r}}^{c}+3, {M_{r}}^{s}<{M_{r}}^{c}+4$.
The blue dotted line is a fitting line with a slope of -1.05. The radial 
distributions for dark matter (cyan solid) and subhalos (purple solid) 
are also shown in the third panel. For clarity, error bars are plotted 
only for four subsamples, and those for other subsamples are comparable.
}

\end{center}
\end{figure}

To see how the distribution changes with the luminosities of both the
centrals and the satellites, we first consider three subsamples of
satellites for each central galaxy sample: ${M_{\rm r}}^{\rm s}<{M_{\rm
    r}}^{\rm c}+\Delta M$, where $\Delta M=2,3,4$. We count satellites
from $0.01$ Mpc to $1$ Mpc away from their central galaxies.  We scale
the surface number density by the mean value within the virial
radius, and the distance by the virial radius.  These radial
distributions are displayed in the logarithmic space in Figure
~\ref{fig:radial_dis_rescl}. The errors are estimated by dividing each 
sample of central galaxies in the specific magnitude bin into eight 
subsamples. Central galaxies are randomly grouped to form subsamples. 

For the most luminous sample of central galaxies, the surface density
drops at $\log(r_{\rm p}/r_{\rm vir})<-1.5$ for all the three
satellite samples.  It corresponds to a projected distance of $<16$
kpc, which is even smaller than the radius of galaxies as luminous as
${M_{\rm r}}^{\rm c}=-23$. Therefore, satellites could be cannibalized
by the central one if their separations become comparable to this
scale, so that the projected surface density is lowered.  For central 
galaxies of ${M_{\rm r}}^{\rm c}=-21$, the surface density rises once
satellites are beyond the virial radius.  These results are consistent
with the outer structure of dark matter halos found in pure N-body
simulations \citep[e.g.]{prada06}, where they showed the
overdensity of dark matter around halos of mass $< 5\times 10^{12}\msun$ 
actually rises with the distance at $\approx 2-5r_{\rm vir}$.  The fact 
can be understood, because groups with mass $<5\times 10^{12}\msun$ are 
located along filaments connecting more massive halos, unlike massive 
clusters (or halos) which are formed at density peaks. 
It is amazing to see that the rising amount found for the
groups of ${M_{\rm r}}^{\rm c}=-21$ at $1-3$ $r_{\rm vir}$ is in good
agreement with that found by \cite{prada06}, although the error bar 
is large at $r_{\rm p}\ga r_{\rm vir}$. 

Within the virial radius, the density profiles of the satellites are
well approximated as linear relations in the logarithmic
space. Therefore, as in previous works \citep{chen06,chen08,cosmos},
we fit the radial distribution with a power law, $\frac{\Sigma(r_{\rm
    p})}{\bar{\Sigma}(<r_{\rm vir})} = A(\frac{r_{\rm
    p}}{r_{\rm vir}})^{\alpha}$, using data in the range
$-1.5<\log({r_{\rm p}/r_{\rm vir}})<0$, where $A=(\alpha+2)/2$.  The
best-fit parameter $\alpha$ is listed in Table 1. All the slopes are
close to $-1.0$ with a small scatter.  They are independent of both
the luminosity of central galaxies and that of satellite galaxies.
The median value of $\alpha$ is $-1.05 \pm 0.08$, very close to the
isothermal value. The blue dotted line in each panel of Figure
~\ref{fig:radial_dis_rescl} shows the power law function with 
$\alpha=-1.05$.

It is interesting to compare the satellite distribution with the
subhalo distribution or the dark matter distribution in the dark
matter halos.  The density profile in the dark matter distribution in
halos has been studied in many works
\citep[e.g.][]{navarro97,ghigna00,jing00,klypin01,navarro04}. Among
them, the most widely used is the NFW profile \citep{navarro97}.  In
an NFW profile, the logarithmic slope changes slowly from $-3$ in 
the outer region to approaching $-1$ in the inner region of the halo.
In contrast, subhalos do not follow the same distribution, but have a
much shallower density distribution than the dark matter density
profile \citep{diemand04,gao04,gao12}. Interestingly, these authors
have shown the subhalo radial distributions are nearly independent of
the subhalo mass when the simulation resolution is properly taken into
account. For comparison, we plot typical dark matter and subhalo
density profiles in Figure ~\ref{fig:radial_dis_rescl}, taken from
\cite{han11} after the projection. Around the radius close to the
virial radius, the logarithmic slope of subhalos is about $-1$, very
close to the observed value of satellites, but much flatter than the
dark matter density profile which has a slope $-3$ in 3-dimentional space. 
This is expected
because the subhalos around the virial radius have suffered from
relatively weak tidal stripping of the host halo, and the subhalo mass
has a close correspondence to the luminosity of satellites. In the
inner part of the halo, the subhalos suffer from a stronger tidal
stripping, with the strength increasing as the
distance from the halo center decreases. This leads to a steeper radial
distribution of the satellite galaxies than that of subhalos. Although
the satellite distribution can be qualitatively explained, the unique
non-trivial features of the radial distribution of satellites as shown
in Figure ~\ref{fig:radial_dis_rescl} should provide a quantitative
test for theories of galaxy formation, especially on the processes
of tidal stripping, dynamical friction, and gas ram pressure.

\begin{figure}
\begin{center}
\plotone{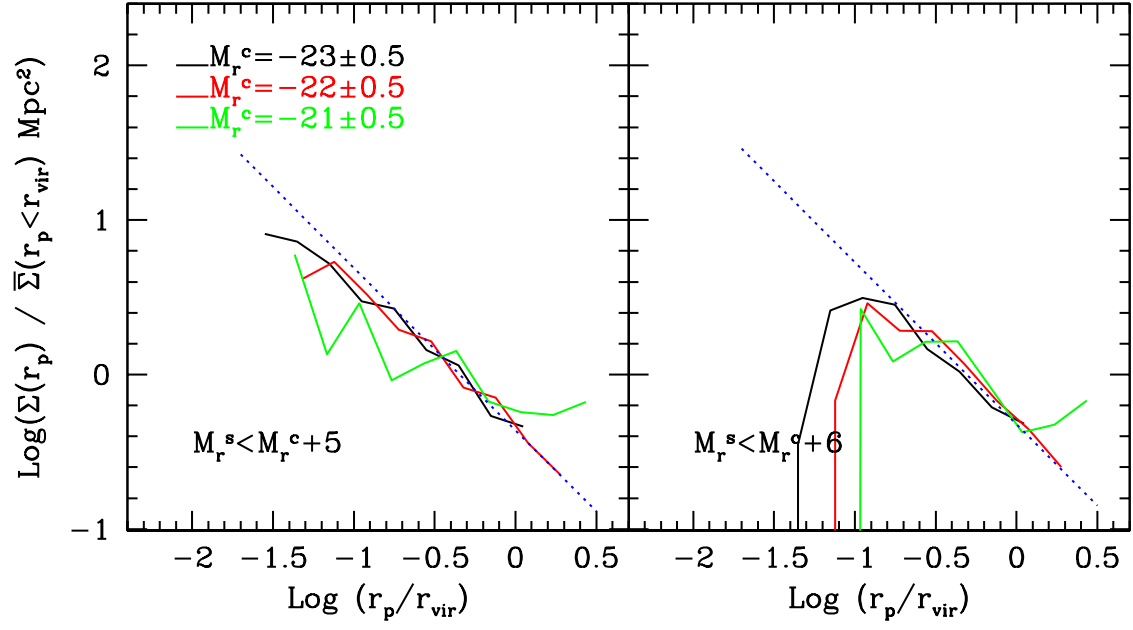}
\caption{\label{fig:radial_dis_rescl_dm56}Similar to Figure ~\ref{fig:radial_dis_rescl}, 
but for satellites with ${M_{r}}^{s}<{M_{r}}^{c}+5$ and ${M_{r}}^{s}<{M_{r}}^{c}+6$.
}
\end{center}
\end{figure}

Next, we consider another two subsamples of fainter satellites:
${M_{\rm r}}^{\rm s}<{M_{\rm r}}^{\rm c}+5$ and ${M_{\rm r}}^{\rm
  s}<{M_{\rm r}}^{\rm c}+6$.  Figure ~\ref{fig:radial_dis_rescl_dm56}
compares radial distributions of these two subsamples with the
distribution function obtained above. The radial distributions for
fainter satellites still conform to the power law form at large
radii. It begins to deviate from the power law at small radii. It
drops sharply for all central samples in the subsample with ${M_{\rm
    r}}^{\rm s}<{M_{\rm r}}^{\rm c}+6$.  The deviation appears at
$r_{\rm p} \sim 0.2 r_{\rm vir}$ for the two
luminous central samples.  For the faintest central sample, the
deviation appears even at a larger distance.

There are two possible reasons why the density of faint satellites
drops at small scales. The first one is observational. Faint
satellites close to a much brighter central galaxy may be missed from
the photometric catalogue. The second reason is physical. In the
central region, subhalos that host satellites are heavily stripped by
the tidal force. This leaves the satellites subject to the tidal
stripping.  The fainter the satellite is, the more it may be stripped.
\cite{cosmos} have examined the faint satellite distribution around
bright centrals by carefully removing the bright central background
with the quality photometric image of the COSMOS field. They found
that the logarithmic slope is $-1.1$ even at the distance close to the
bright central galaxy. With their findings, we think that the drop of
the faint satellite density around the central ones is more like
caused by the observation or the data reduction. One needs to take
this into account when computing the luminosity function and the
abundance of satellites in observations. Considering the fact that the
surface density of the faintest satellites starts to drop at
$r_{\rm p}/r_{\rm vir} \sim 0.2$ for the $M_{\rm r}=-21$
central sample, we estimate that both observational quantities (the
luminosity function and the abundance) could be underestimated by 20
percent. This effect is small for other cases of satellites and
central galaxies.

\subsection{Projected two-point cross correlation function}

Based on the power law radial distribution function obtained above, 
we obtain a smoothed
surface number density $\Sigma(r_{\rm p})$ of satellite galaxies.
Then we adopt a method similar to \cite{wenting11} to calculate the
projected two-point cross correlation function $w_{\rm p}(r_{\rm p})$
of the central galaxies and the satellites, based on the relation 
$\Sigma(r_{\rm p})=w_{\rm p}(r_{\rm p})n_{\rm g}$.
We estimate $n_{\rm g}$ from
the $r^{\prime}$-band luminosity function, which is approximated by
that of the SDSS $r^{0.1}$-band given in \cite{blanton03}.

\begin{figure}
\begin{center}
\plotone{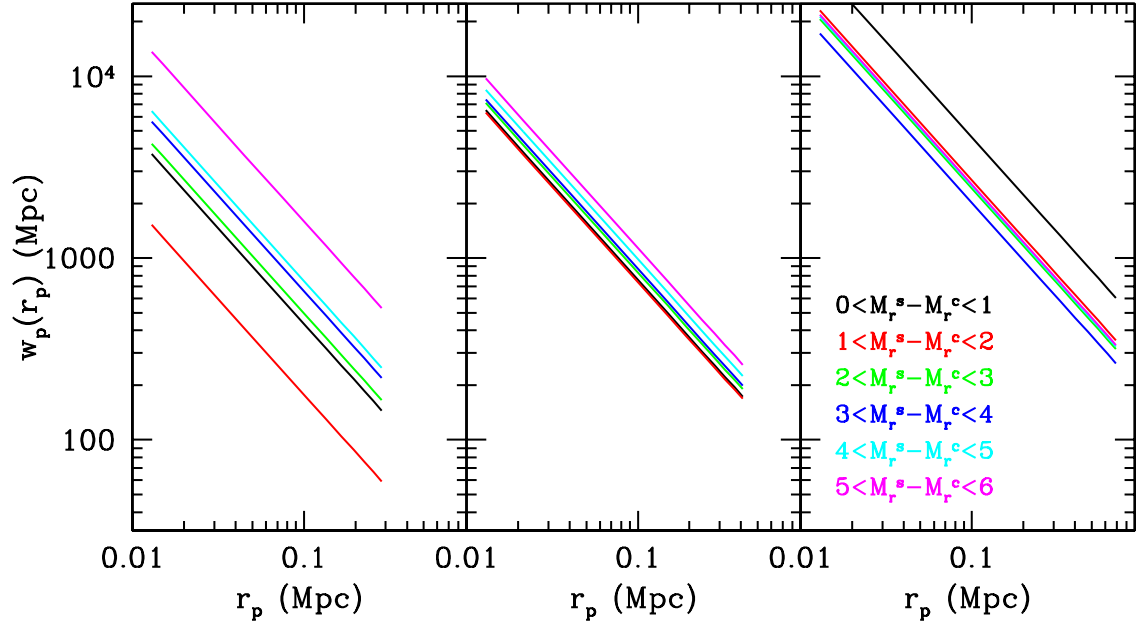}
\caption{\label{fig:wp}Projected correlation functions for central galaxies 
with ${M_{r}}^{c}=-21,-22,-23$ from left to right, extending to virial radii.
}
\end{center}
\end{figure}

Figure \ref{fig:wp} shows the projected cross correlation functions
between central galaxies in three magnitude bins and satellites in six
relative magnitude bins. The logarithmic slope is the same for all
cross correlation functions, as expected from the number density
profile.  The amplitude of $w_{\rm p}$ varies both with the luminosity
of central galaxies and satellite galaxies. A conspicuous feature
displayed in the plots is that, the clustering amplitudes do not
exhibit a monotonic dependence on satellite luminosity.  This
is most pronounced for the faintest central magnitude bin. The
amplitude first decreases as the satellite luminosity is lowered to
${M_{\rm r}}^{\rm s}-{M_{\rm r}}^{\rm c}=2$, and then it increases as
the satellite luminosity continues to be lowered. The turning point is
relatively at larger $\Delta M={M_{\rm r}}^{\rm s}-{M_{\rm r}}^{\rm
  c}$ in the most luminous central sample. This happens at ${M_{\rm
    r}}^{\rm s}-{M_{\rm r}}^{\rm c}=4$.  The relatively close
amplitudes between different satellite bins for central galaxies with
${M_{\rm r}}^{\rm c}=-22,-23$ are generally consistent with those found
in \cite{wenting11}. Their sample is not deep enough, however, so that
the non-monotonic trend cannot be detected in their work.  A similar
behavior was found in \cite{li06} and \cite{zehavi11}. 
\cite{li06} found that, the projected auto-correlation function
for the red galaxies at $r_{\rm p}=0.2h^{-1}$ Mpc reaches the lowest
value when the galaxy luminosity is around $-20.5+5\log10(h)$. 
\cite{zehavi11} showed with their sample
of red galaxies that, the decreasing trend of projected
auto-correlation functions with luminosities is reversed at $r_{\rm
  p}<\sim 2h^{-1}$ Mpc when the galaxy luminosity is fainter than
$-20+5\log10(h)$.  Since the cross correlation is an approximation of
the geometric mean of the two auto-correlations \citep{szapudi92}, the
rebound behavior in the auto-correlation functions indicates a similar
feature in the cross correlations.  The higher clustering amplitudes
of faint satellite galaxies (especially those with ${M_{\rm r}}^{\rm
  s}-{M_{\rm r}}^{\rm c}\sim 5.5$) around their centrals suggests a
higher fraction of these galaxies exists as satellites in halos being
considered than the more luminous galaxies. 

\section{Luminosity Functions of satellites}

\begin{figure}
\begin{center}
\plotone{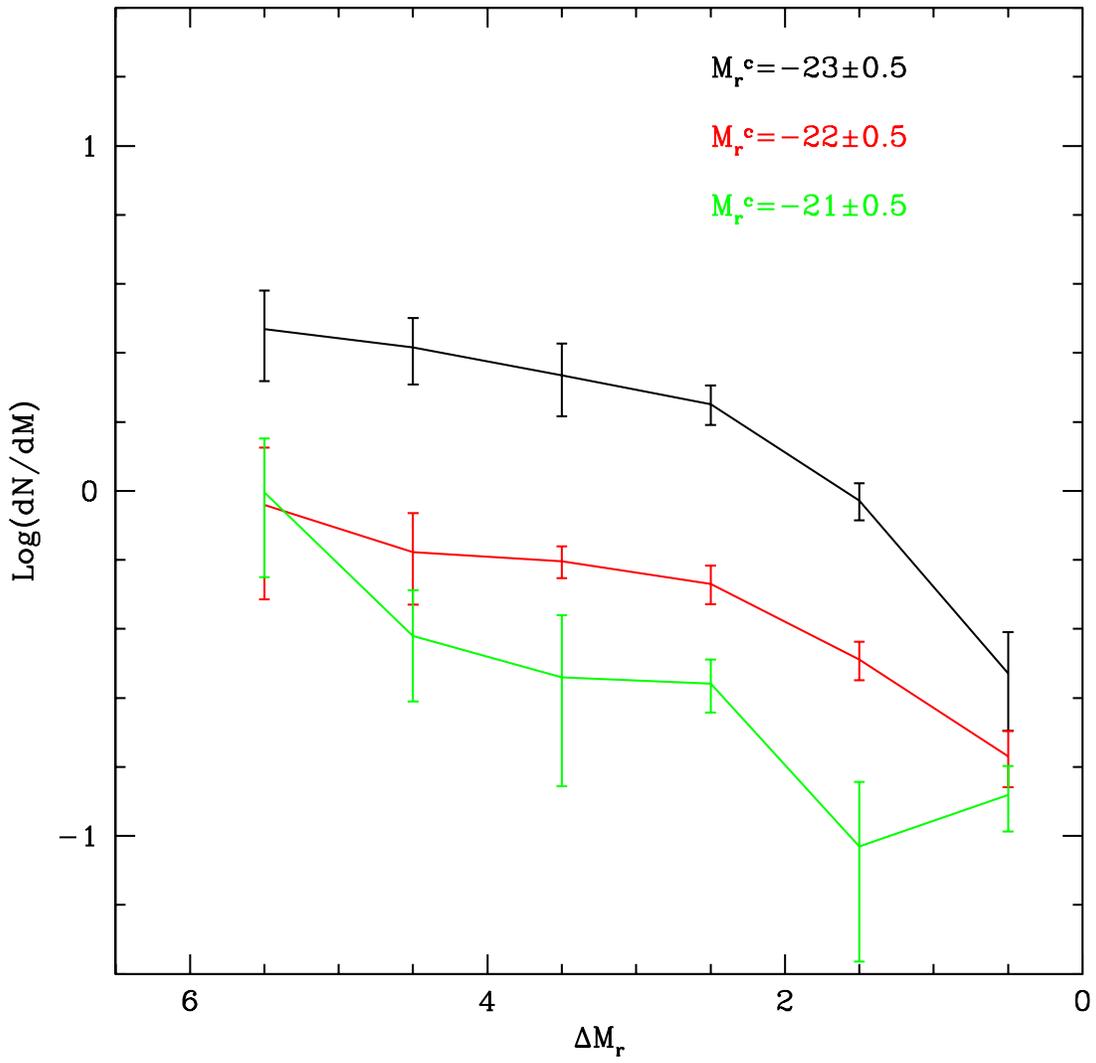}
\caption{\label{fig:lf_rvir}Luminosity functions for central galaxies in the 
three magnitude bins within virial radii.
}
\end{center}
\end{figure}

Figure ~\ref{fig:lf_rvir} shows the luminosity functions of satellites
for central galaxies in the three magnitude bins: ${M_{\rm r}}^{\rm
  c}=-21,-22$, and $-23$. The number of satellites per magnitude for
each central galaxy is plotted against the $r^{\prime}$ band magnitude
difference between satellites and their central galaxies, $\Delta
M={M_{\rm r}}^{\rm s}-{M_{\rm r}}^{\rm c}$. As in section \ref{sec:radial_dis}, 
we estimate the errors from eight randomly grouped subsamples. 
Satellites are counted within the virial radius of their
host halo, and the number count plotted in this figure and the
following figures are $N_i$ (see Equation ~\ref{eq:Ni}).

With Figure ~\ref{fig:lf_rvir}, we compare the population of
satellites with that of subhalos to find their connections.  Using the
scaled subhalo mass which is the subhalo mass divided by their host
halo mass, $M_{\rm sub}/M_{\rm host}$, studies based on N-body
simulations have found that the subhalo mass function
$dN/d\log \frac{M_{\rm sub}}{M_{\rm host}}\propto (\frac{M_{\rm sub}}
{M_{\rm host}})^{\beta} $ with $\beta\approx -0.9$, depending on the host halo
mass very weakly (at most ${M_{\rm host}}^{0.1}$, \citealp{angulo09,han11}). 
If the mass-to-light ratio
is the same for all (sub)halos, we can easily deduce that the number
of satellites per central galaxy when expressed as a function of
$\Delta M$ as in Figure ~\ref{fig:lf_rvir} should also depend little
on the luminosity of the centrals, but increase linearly with $\Delta
M$.  In contrast, Figure ~\ref{fig:lf_rvir} demonstrates a different
picture. The major contribution of the difference between the three
magnitude bins comes from a varying mass-to-light ratio with halo
mass.  We know that, the mass-to-light ratio at $z\sim 0$ is the
lowest at $M_{\rm halo} \sim 10^{12}\msun$
\citep{yang03,zheng07,zehavi11}, which marks the halo mass with the
highest star formation efficiency. Above this mass, the mass-to-light
ratio rises and the light increases more and more slowly with
mass. Therefore, the number of satellites at fixed $\Delta M$ is
different for central galaxies with different luminosities. It depends
on the difference in the mass-to-light ratio between the central and
the satellite galaxies.  This difference in mass-to-light ratio
does not solely come from the difference in star formation efficiency,
but also from the fact that satellite halos are stripped of their mass after
being accreted to host halos, leading to a lower mass-to-light
ratio compared to that at accretion time.

\begin{figure}
\begin{center}
\plotone{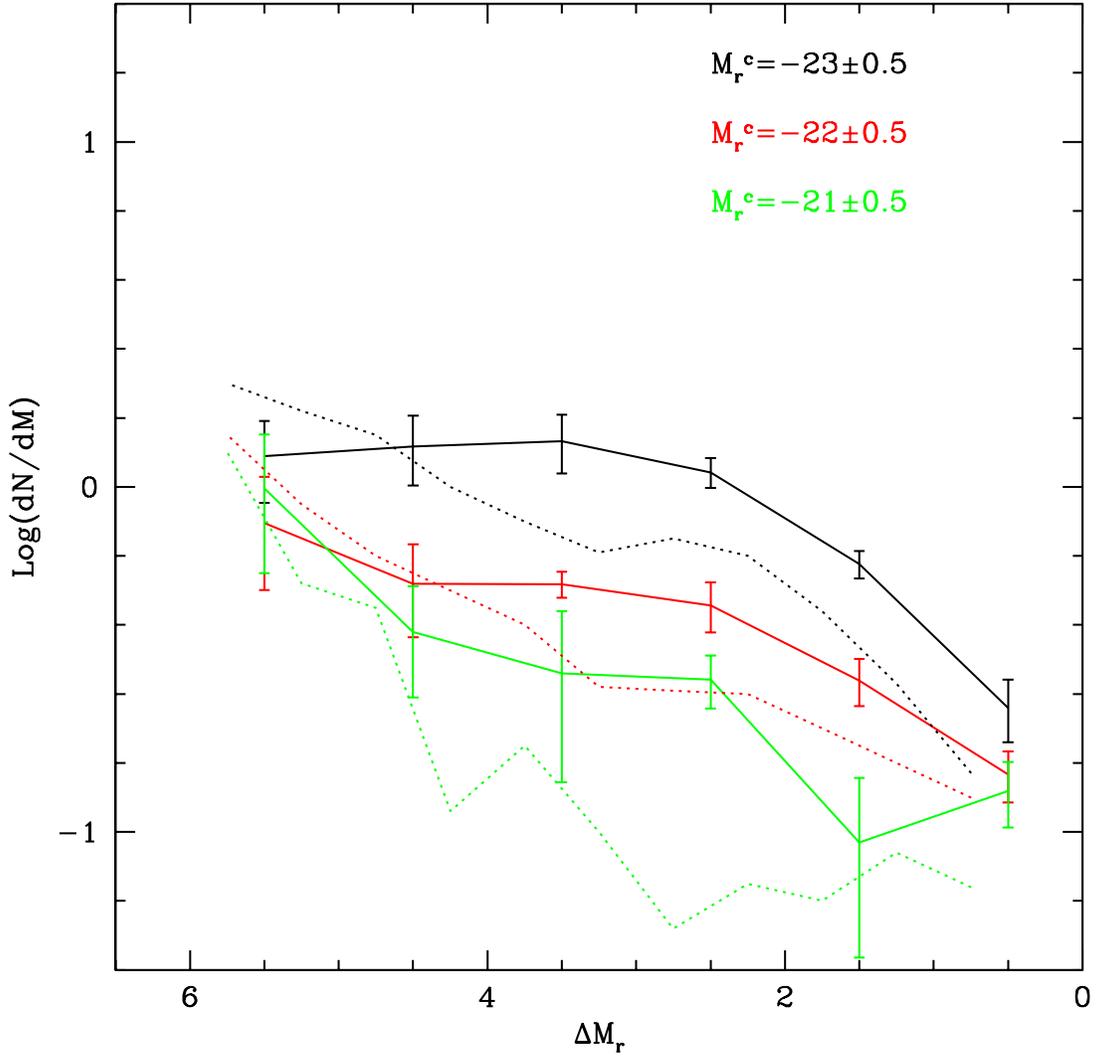}
\caption{\label{fig:lf}Similar to Figure ~\ref{fig:lf_rvir}, but satellites 
are counted within projected distance 0.3 Mpc. In comparison, results of 
\cite{guo11} are shown with dotted lines.
}
\end{center}
\end{figure}

To compare with a previous work by \cite{guo11}, we also calculate the
luminosity functions within $0.3$ Mpc of central galaxies (solid lines
in Figure ~\ref{fig:lf}). Results from \cite{guo11} are shown with
dotted lines. We see that, our results are higher than theirs in all
luminosity bins.  Although the $r^{\prime}$ filter we use is slightly
different from the SDSS $r$ filter, according to the color
transformations with the SDSS filters \citep{regnault09}, we find a
tiny difference of $0.01$ mag between $r^{\prime}$ and $r_{\rm SDSS}$ 
for central galaxies we consider here.
One reason for the discrepancy is the different method adopted in the
subtraction of the background galaxies.  \cite{guo11} take galaxies
that are 0.3 Mpc to 0.6 Mpc away from their central galaxies as a
proxy for background galaxies. However, this distance range is well
within the virial radius for galaxies of ${M_{\rm r}}^{\rm c}=-23$,
and partly within the virial radius for galaxies of ${M_{\rm r}}^{\rm
  c}=-22$, according to the virial radius we calculated
above. Therefore, the background may be over-subtracted for these two
luminosity bins.  For central galaxies of ${M_{\rm r}}^{\rm c}=-21$,
although the radial range of 0.3-0.6 Mpc already falls out of the
virial radius, it is however not guaranteed that the background
galaxies are properly subtracted in this way. In fact, as shown by
Figure \ref{fig:radial_dis_rescl} and the N-body results of
\cite{prada06}, the nearby structures outside the virial
radius of the host halo of a Milky Way-like galaxy have elevated the 
surrounding galaxy density. We also note that in their work, the 
correlated galaxies outside the virial radius along the line of sight 
are not explicitly subtracted.

\begin{figure}
\begin{center}
\plotone{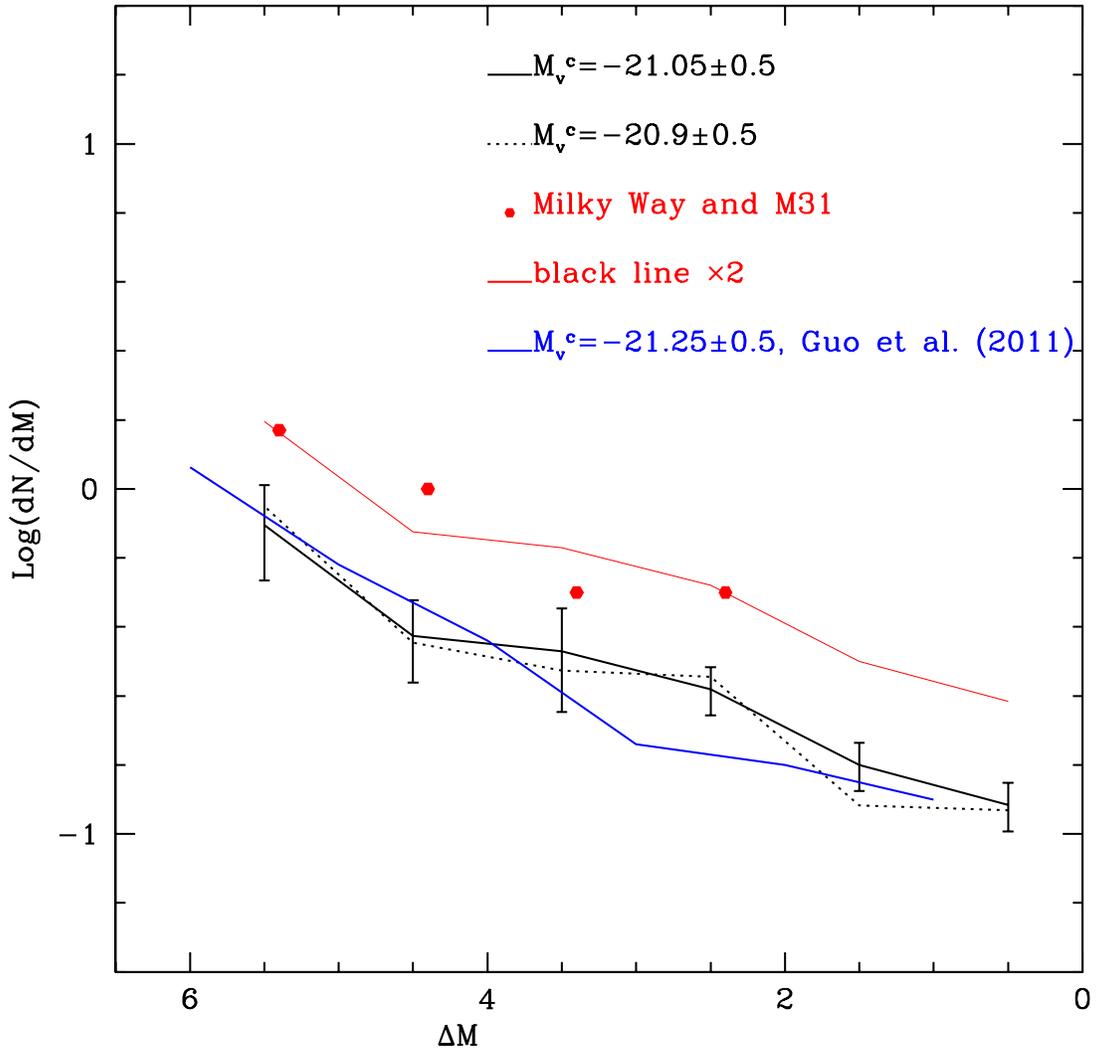}
\caption{\label{fig:lf_milkyway}Luminosity functions for Milky Way-like 
central galaxies ($M_{\rm v}=-20.9 \pm 0.5$) (black dotted) and Milky 
Way/M31-like central galaxies ($M_{\rm v}=-21.05 \pm 0.5$) (black solid).
The luminosity function of the Milky Way and M31 is plotted with red points 
\citep{koposov08}. The results of \cite{guo11} are shown with the blue 
solid line.
}
\end{center}
\end{figure}

Next, we study the satellite abundance in MW/M31-like galaxies. The
V-band magnitude (vega system) of the Milky Way and M31 is $-20.9$ and
$-21.2$ respectively \citep{vandenBergh00}, which gives a mean
magnitude of $-21.05$ for the MW/M31 system. For galaxies with $M_{\rm
  v} = -20.9 \pm 0.5$ (the conversion from vega magnitude to AB
magnitude is ignored, which is only $0.02$), the mean magnitude in the
$r^{\prime}$ band is $M_{\rm r}=-21.3$.  \cite{liu11} 
discussed that $M_{\rm v} = -20.9 \pm 0.5$ corresponds to ${M_{\rm
    r}}^{0.1}=-21.2$, in which ${M_{\rm r}}^{0.1}$ is the SDSS r band
magnitude which is $k$-corrected to $z=0.1$. Considering the
$k$-correction from $z=0.1$ to $z=0.0$ is of order $0.1$, the two
values are consistent.  In addition, for galaxies with $M_{\rm v} = -21.05 \pm
0.5$, the mean $M_{\rm r}=-21.4$. We consider both samples, one with
$M_{\rm v} = -20.9 \pm 0.5$, and the other with $M_{\rm v} = -21.05
\pm 0.5$.  We convert the $r^{\prime}$ band to the $V$ band magnitude
with the $V-r^{\prime}$ color.  
Then we search for the centrals in the $ M_{\rm v}$ interval from all
the matched central galaxies.  The difference of the two samples are
very small, as shown in Figure ~\ref{fig:lf_milkyway}.  We can see from
Figure ~\ref{fig:lf_milkyway} that, the luminosity function of bright
satellites in the Milky Way and M31 (red points, \citealt{koposov08})
lies above that of our MW/M31-like galaxies by a factor of $2$.  A
comparison with the results in \cite{guo11} is also given in this
figure.  It shows a reasonable agreement except that the amplitude
around $\Delta M=3$ is a bit lower in their results, despite the
different methods used. We also note that the satellite luminosity
functions of the two Milky Way like galaxy samples ($M_{\rm v} =
-21.05$ and $M_{\rm v} = -20.9$) are close to that of ${M_{\rm
    r}}^{\rm c}=-21$ galaxies in Figure ~\ref{fig:lf} in our analysis
as expected, but they are not apparently in \cite{guo11} (comparing
the green dotted line in Figure ~\ref{fig:lf} with the blue line Figure
~\ref{fig:lf_milkyway}).

\section{Conclusions}

In this work, we have investigated the distributions of satellites
around their central galaxies using the photometric catalogue of the
CFHT Legacy Survey. We use the SDSS/DR7 group catalog of \cite{yang07}
to identify the central galaxies. The method we have used can count in 
all the candidate satellites while subtracting the background galaxies
accurately. We have focused on three samples of central galaxies with
magnitudes of ${M_{\rm r}}^{\rm c}=-21,-22$, and $-23$. Our
main results can be summarized as follows.
\begin{itemize}
\item Over a wide radial range $-1.5<\log({r_{\rm p}/r_{\rm vir}})<0$,
the distribution of projected radial number density obeys a power law
form with the best-fit logarithmic slope of $-1.05$, indicating that the
3-dimensional number density follows approximately an isothermal
distribution. The radial number density profile, if scaled by the mean
number density within the virial radius, is independent of the
luminosity of central galaxies and satellite galaxies.

\item The projected cross correlation functions between central galaxies and
their satellites exhibit a non-monotonic trend with satellite
luminosity. As the satellite luminosity decreases, the amplitude of
the cross-correlation function first decreases, before it increases
after the satellite luminosity crosses a turning point. This is most
pronounced for central galaxies with ${M_{\rm r}}^{\rm c}=-21$, where
the decreasing trend of the clustering amplitude as the satellite
luminosity decreases is reversed when satellites are fainter than
central galaxies by more than 2 magnitudes.

\item The Milky Way/M31 system has about twice as many satellites with
magnitude difference $\Delta M={M_{\rm r}}^{\rm s}-{M_{\rm r}}^{\rm
  c}<6$ as we have obtained statistically for galaxies of the same
luminosity. This over-abundance shows that the MW/M31 is atypical of
central galaxies with the same luminosity in the distribution of
satellites.
\end{itemize}

\acknowledgments
This work is  sponsored  by NSFC  (11121062, 10878001, 11033006, 
11003035, 11173045, 11233005) and the CAS/SAFEA International Partnership 
Program for  Creative Research Teams  (KJCX2-YW-T23).
CL acknowledges the support of the 100-Talent Program of Chinese 
Academy of Sciences (CAS), Shanghai  Pujiang Programme (no. 11PJ1411600)
and the exchange program between Max Planck Society and CAS.

We would like to thank Xiaohu Yang for providing us the SDSS/DR7 group
catalog. We are grateful to Stephen Gwyn for making the photometric 
catalog of CFHTLS publicly available. The catalog is based on observations
obtained with MegaPrime/MegaCam, a joint project of CFHT and CEA/DAPNIA, 
at the Canada-France-Hawaii Telescope (CFHT) which is operated by the 
National Research Council (NRC) of Canada, the Institut National des 
Science de l'Univers of the Centre National de la Recherche Scientifique 
(CNRS) of France, and the University of Hawaii. This work is based in 
part on data products produced at the Canadian Astronomy Data Centre 
as part of the Canada-France-Hawaii Telescope Legacy Survey, a 
collaborative project of NRC and CNRS.

Funding for  the SDSS and SDSS-II  has been provided by  the Alfred P.
Sloan Foundation, the Participating Institutions, the National Science
Foundation, the  U.S.  Department of Energy,  the National Aeronautics
and Space Administration, the  Japanese Monbukagakusho, the Max Planck
Society,  and the Higher  Education Funding  Council for  England. The
SDSS Web  Site is  http://www.sdss.org/.  The SDSS  is managed  by the
Astrophysical    Research    Consortium    for    the    Participating
Institutions. The  Participating Institutions are  the American Museum
of  Natural History,  Astrophysical Institute  Potsdam,  University of
Basel,  University  of  Cambridge,  Case Western  Reserve  University,
University of Chicago, Drexel  University, Fermilab, the Institute for
Advanced   Study,  the  Japan   Participation  Group,   Johns  Hopkins
University, the  Joint Institute  for Nuclear Astrophysics,  the Kavli
Institute  for   Particle  Astrophysics  and   Cosmology,  the  Korean
Scientist Group, the Chinese  Academy of Sciences (LAMOST), Los Alamos
National  Laboratory, the  Max-Planck-Institute for  Astronomy (MPIA),
the  Max-Planck-Institute  for Astrophysics  (MPA),  New Mexico  State
University,   Ohio  State   University,   University  of   Pittsburgh,
University  of  Portsmouth, Princeton  University,  the United  States
Naval Observatory, and the University of Washington.

\begin{table}
\caption{Best-Fit Power Laws of Scaled Radial Distributions}
\vspace*{7pt}
\centering
\begin{tabular}{c c c c}
\hline\hline

\multicolumn{2}{c}{Sample} & $\alpha$ & $\chi2/dof$ \\[0.5ex]
\hline
  &${M_{r}}^{s}<{M_{r}}^{c}+2$ & $-1.08\pm0.09$ &  0.64 \\[-2ex]
  \raisebox{-2.0ex}{${M_{r}}^{c}=-23\pm 0.5$}
  &\raisebox{-2.0ex}{${M_{r}}^{s}<{M_{r}}^{c}+3$} &\raisebox{-2.0ex}{$-1.10\pm0.07$} & \raisebox{-2.0ex}{1.50} \\[2ex]
  &${M_{r}}^{s}<{M_{r}}^{c}+4$ & $-1.02\pm0.08$ & 1.09 \\[2ex]

  &${M_{r}}^{s}<{M_{r}}^{c}+2$ & $-1.04\pm0.08$ & 1.31 \\[-2ex]
  \raisebox{-2.0ex}{${M_{r}}^{c}=-22\pm 0.5$}
  &\raisebox{-2.0ex}{${M_{r}}^{s}<{M_{r}}^{c}+3$} &\raisebox{-2.0ex}{$-0.94\pm0.04$} & \raisebox{-2.0ex}{0.69} \\[2ex]
  &${M_{r}}^{s}<{M_{r}}^{c}+4$ & $-1.07\pm0.05$ & 1.12 \\[2ex]

  &${M_{r}}^{s}<{M_{r}}^{c}+2$ & $-1.08\pm0.32$ & 0.53 \\[-2ex]
  \raisebox{-2.0ex}{${M_{r}}^{c}=-21\pm 0.5$}
  &\raisebox{-2.0ex}{${M_{r}}^{s}<{M_{r}}^{c}+3$} &\raisebox{-2.0ex}{$-1.01\pm0.16$} & \raisebox{-2.0ex}{1.54} \\[2ex]
  &${M_{r}}^{s}<{M_{r}}^{c}+4$ &$-1.05\pm0.15$ & 0.46 \\[2ex]

\hline
\end{tabular}
\label{tab:fitting}
\end{table}

\begin{thebibliography}{}
\bibitem[Abazajian et al.(2009)]{abazajian09} Abazajian, K.~N., 
Adelman-McCarthy, J.~K., Ag{\"u}eros, M.~A., et al.\ 2009, \apjs, 182, 543 


\bibitem[Angulo et al.(2009)]{angulo09} Angulo, R.~E., Lacey, 
C.~G., Baugh, C.~M., \& Frenk, C.~S.\ 2009, \mnras, 399, 983 


\bibitem[Arnouts et al.(1999)]{arnouts99} Arnouts, S., Cristiani, 
S., Moscardini, L., et al.\ 1999, \mnras, 310, 540 


\bibitem[Blanton et al.(2003)]{blanton03} Blanton, M.~R., Hogg, 
D.~W., Bahcall, N.~A., et al.\ 2003, \apj, 592, 819 


\bibitem[Blanton et al.(2005)]{blanton05} Blanton, M.~R., 
Schlegel, D.~J., Strauss, M.~A., et al.\ 2005, \aj, 129, 2562 


\bibitem[Boylan-Kolchin et al.(2010)]{boylan-kolchin10} Boylan-Kolchin, 
M., Springel, V., White, S.~D.~M., \& Jenkins, A.\ 2010, \mnras, 406, 896 


\bibitem[Busha et al.(2011)]{busha11} Busha, M.~T., Wechsler, 
R.~H., Behroozi, P.~S., et al.\ 2011, \apj, 743, 117 


\bibitem[Chen(2008)]{chen08} Chen, J.\ 2008, \aap, 484, 347 


\bibitem[Chen et al.(2006)]{chen06} Chen, J., Kravtsov, A.~V., 
Prada, F., et al.\ 2006, \apj, 647, 86 


\bibitem[Coupon et 
al.(2009)]{coupon09} Coupon, J., Ilbert, O., Kilbinger, M., et al.\ 2009, \aap, 500, 981 


\bibitem[Davis 
\& Peebles(1983)]{davis83} Davis, M., \& Peebles, P.~J.~E.\ 1983, \apj, 267, 465 


\bibitem[Dehnen et al.(2006)]{dehnen06} Dehnen, W., McLaughlin, 
D.~E., \& Sachania, J.\ 2006, \mnras, 369, 1688 


\bibitem[Diemand et al.(2004)]{diemand04} Diemand, J., Moore, B., 
\& Stadel, J.\ 2004, \mnras, 352, 535 


\bibitem[Gao et al.(2012)]{gao12} Gao, L., Navarro, J.~F., 
Frenk, C.~S., et al.\ 2012, arXiv:1201.1940 


\bibitem[Gao et al.(2004)]{gao04} Gao, L., White, S.~D.~M., 
Jenkins, A., Stoehr, F., \& Springel, V.\ 2004, \mnras, 355, 819 


\bibitem[Gao et al.(2008)]{gao08} Gao, L., Navarro, J.~F., 
Cole, S., et al.\ 2008, \mnras, 387, 536 


\bibitem[Ghigna et al.(2000)]{ghigna00} Ghigna, S., Moore, B., 
Governato, F., et al.\ 2000, \apj, 544, 616 


\bibitem[Guo et al.(2010)]{guoqi10} Guo, Q., White, S., Li, C.,
\& Boylan-Kolchin, M.\ 2010, \mnras, 404, 1111


\bibitem[Guo et al.(2011)]{guo11} Guo, Q., Cole, S., Eke, V., 
\& Frenk, C.\ 2011, \mnras, 417, 370 


\bibitem[Gwyn(2011)]{gwyn11} Gwyn, S.~D.~J.\ 2011, 
arXiv:1101.1084 


\bibitem[Han et al.(2011)]{han11} Han, J., Jing, Y.~P., Wang, 
H., \& Wang, W.\ 2011, arXiv:1103.2099 


\bibitem[Ilbert et 
al.(2006)]{ilbert06} Ilbert, O., Arnouts, S., McCracken, H.~J., et al.\ 2006, \aap, 457, 841


\bibitem[Jiang et al.(2008)]{jiang08} Jiang, C.~Y., Jing, 
Y.~P., Faltenbacher, A., Lin, W.~P., \& Li, C.\ 2008, \apj, 675, 1095 


\bibitem[Jing 
\& Suto(2000)]{jing00} Jing, Y.~P., \& Suto, Y.\ 2000, \apjl, 529, L69 


\bibitem[Klypin et al.(2001)]{klypin01} Klypin, A., Kravtsov, 
A.~V., Bullock, J.~S., \& Primack, J.~R.\ 2001, \apj, 554, 903 


\bibitem[Koposov et al.(2008)]{koposov08} Koposov, S., Belokurov, 
V., Evans, N.~W., et al.\ 2008, \apj, 686, 279 


\bibitem[Koposov et al.(2009)]{koposov09} Koposov, S.~E., Yoo, 
J., Rix, H.-W., et al.\ 2009, \apj, 696, 2179 


\bibitem[Kravtsov(2010)]{kravtsov10} Kravtsov, A.\ 2010, Advances 
in Astronomy, 2010,  


\bibitem[Kravtsov et al.(2004)]{kravtsov04} Kravtsov, A.~V., 
Gnedin, O.~Y., \& Klypin, A.~A.\ 2004, \apj, 609, 482 


\bibitem[Lares et al.(2011)]{lares11} Lares, M., Lambas, D.~G., 
\& Dom{\'{\i}}nguez, M.~J.\ 2011, \aj, 142, 13 


\bibitem[Li et al.(2012)]{li12} Li, C., Jing, Y.~P., Mao,
S., et al.\ 2012, arXiv:1206.3566


\bibitem[Li et al.(2006)]{li06} Li, C., Kauffmann, G., Jing, 
Y.~P., et al.\ 2006, \mnras, 368, 21 


\bibitem[Liu et al.(2011)]{liu11} Liu, L., Gerke, B.~F., 
Wechsler, R.~H., Behroozi, P.~S., \& Busha, M.~T.\ 2011, \apj, 733, 62 


\bibitem[Lovell et al.(2012)]{lovell12} Lovell, M.~R., Eke, V., 
Frenk, C.~S., et al.\ 2012, \mnras, 420, 2318 


\bibitem[Ludlow et al.(2009)]{ludlow09} Ludlow, A.~D., Navarro, 
J.~F., Springel, V., et al.\ 2009, \apj, 692, 931 


\bibitem[Madau et al.(2008)]{madau08} Madau, P., Diemand, J., 
\& Kuhlen, M.\ 2008, \apj, 679, 1260 


\bibitem[Madore et al.(2004)]{madore04} Madore, B.~F., Freedman, 
W.~L., \& Bothun, G.~D.\ 2004, \apj, 607, 810 


\bibitem[Mandelbaum et al.(2006)]{mandelbaum06} Mandelbaum, R., 
Seljak, U., Kauffmann, G., Hirata, C.~M., 
\& Brinkmann, J.\ 2006, \mnras, 368, 715


\bibitem[Merritt et al.(2006)]{merritt06} Merritt, D., Graham, 
A.~W., Moore, B., Diemand, J., \& Terzi{\'c}, B.\ 2006, \aj, 132, 2685 


\bibitem[Merritt et al.(2005)]{merritt05} Merritt, D., Navarro, 
J.~F., Ludlow, A., \& Jenkins, A.\ 2005, \apjl, 624, L85 


\bibitem[More et al.(2011)]{more11} More, S., van den Bosch,
F.~C., Cacciato, M., et al.\ 2011, \mnras, 410, 210


\bibitem[Moster et al.(2010)]{moster10} Moster, B.~P.,
Somerville, R.~S., Maulbetsch, C., et al.\ 2010, \apj, 710, 903


\bibitem[Navarro et al.(2004)]{navarro04} Navarro, J.~F., 
Hayashi, E., Power, C., et al.\ 2004, \mnras, 349, 1039 


\bibitem[Navarro et al.(1997)]{navarro97} Navarro, J.~F., Frenk, 
C.~S., \& White, S.~D.~M.\ 1997, \apj, 490, 493 


\bibitem[Nierenberg et al.(2012)]{cosmos} Nierenberg, A.~M., 
Auger, M.~W., Treu, T., et al.\ 2012, arXiv:1202.2125 


\bibitem[Prada et al.(2006)]{prada06} Prada, F., Klypin, A.~A., 
Simonneau, E., et al.\ 2006, \apj, 645, 1001 


\bibitem[Regnault et 
al.(2009)]{regnault09} Regnault, N., Conley, A., Guy, J., et al.\ 2009, \aap, 506, 999 


\bibitem[Robotham et al.(2012)]{robotham12} Robotham, A.~S.~G., 
Baldry, I.~K., Bland-Hawthorn, J., et al.\ 2012, \mnras, 424, 1448 


\bibitem[Sakamoto et 
al.(2003)]{sakamoto03} Sakamoto, T., Chiba, M., \& Beers, T.~C.\ 2003, \aap, 397, 899 


\bibitem[Sales
\& Lambas(2005)]{sales05} Sales, L., \& Lambas, D.~G.\ 2005, \mnras, 356, 1045


\bibitem[Smith et al.(2004)]{smith04} Smith, R.~M., 
Mart{\'{\i}}nez, V.~J., \& Graham, M.~J.\ 2004, \apj, 617, 1017 


\bibitem[Stanimirovi{\'c} et al.(2004)]{stanimirovic04} 
Stanimirovi{\'c}, S., Staveley-Smith, L., 
\& Jones, P.~A.\ 2004, \apj, 604, 176 


\bibitem[Szapudi et al.(1992)]{szapudi92} Szapudi, I., Szalay, 
A.~S., \& Boschan, P.\ 1992, \apj, 390, 350 


\bibitem[Tinker et al.(2005)]{tinker05} Tinker, J.~L., Weinberg, 
D.~H., Zheng, Z., \& Zehavi, I.\ 2005, \apj, 631, 41 


\bibitem[van den Bergh(2000)]{vandenBergh00} van den Bergh, S.\ 2000, 
\pasp, 112, 529


\bibitem[van der Marel et al.(2002)]{vanderMarel02} van der Marel, 
R.~P., Alves, D.~R., Hardy, E., \& Suntzeff, N.~B.\ 2002, \aj, 124, 2639 


\bibitem[Wang et al.(2012)]{wang12} Wang, J., Frenk, C.~S.,
Navarro, J.~F., \& Gao, L.\ 2012, arXiv:1203.4097


\bibitem[Wang 
\& Jing(2010)]{wanglan10} Wang, L., \& Jing, Y.~P.\ 2010, \mnras, 402, 1796 


\bibitem[Wang et al.(2006)]{wanglan06} Wang, L., Li, C., 
Kauffmann, G., \& De Lucia, G.\ 2006, \mnras, 371, 537


\bibitem[Wang et al.(2011)]{wenting11} Wang, W., Jing, Y.~P., Li, 
C., Okumura, T., \& Han, J.\ 2011, \apj, 734, 88 


\bibitem[Watson et al.(2010)]{watson10} Watson, D.~F., Berlind, 
A.~A., McBride, C.~K., \& Masjedi, M.\ 2010, \apj, 709, 115 


\bibitem[Wilkinson 
\& Evans(1999)]{wilkinson99} Wilkinson, M.~I., \& Evans, N.~W.\ 1999, \mnras, 310, 645


\bibitem[Xue et al.(2008)]{xue08} Xue, X.~X., Rix, H.~W., 
Zhao, G., et al.\ 2008, \apj, 684, 1143


\bibitem[Yang et al.(2003)]{yang03} Yang, X., Mo, H.~J., 
\& van den Bosch, F.~C.\ 2003, \mnras, 339, 1057 


\bibitem[Yang et al.(2007)]{yang07} Yang, X., Mo, H.~J., van 
den Bosch, F.~C., et al.\ 2007, \apj, 671, 153 


\bibitem[York et al.(2000)]{york00} York, D.~G., Adelman, J., 
Anderson, J.~E., Jr., et al.\ 2000, \aj, 120, 1579 


\bibitem[Zavala et al.(2009)]{zavala09} Zavala, J., Jing, Y.~P., 
Faltenbacher, A., et al.\ 2009, \apj, 700, 1779 


\bibitem[Zehavi et al.(2005)]{zehavi05} Zehavi, I., Zheng, Z., 
Weinberg, D.~H., et al.\ 2005, \apj, 630, 1 


\bibitem[Zehavi et al.(2011)]{zehavi11} Zehavi, I., Zheng, Z., 
Weinberg, D.~H., et al.\ 2011, \apj, 736, 59 


\bibitem[Zheng et al.(2007)]{zheng07} Zheng, Z., Coil, A.~L., 
\& Zehavi, I.\ 2007, \apj, 667, 760 



\end{thebibliography}
\end{document}